Conference Proceedings

# Towards the Development of Detection of Learned Helplessness in Mathematics: Design and Data Collection Challenges from a Developing Country Perspective

John Paul P. Miranda[1,2]*  , Rex P. Bringula[2] , Laharni S. Simpao[1] , Jordan L. Salenga[1] ,
Juvy C. Grume[1] , Madilaine Claire B. Nacianceno[1]  , Lester G. Loyola[3] , Jaymark A. Yambao[1]

1. **Don Honorio Ventura State University**, Pampanga, Philippines
2. **University of the East**, Manila, Philippines
3. **National University, Philippines**, Manila, Philippines

**\* Correspondence:**
John Paul P. Miranda, Don Honorio Ventura State University, jppmiranda@dhvsu.edu.ph

## ABSTRACT

This study investigates the challenges in designing, data collection, and implementation of a web-based Tutoring System (TS) for teaching linear equations within a developing country context. Originally designed as an Android app, the system was redeveloped as a web application to facilitate cross-platform access and data collection. This redesign enabled enhanced tracking through interaction logs and included features like problem skipping, hints, difficulty-based problem sequencing, and game modes with adaptable progression (e.g., easy-to-hard, hard-to-easy). The main objective was to document the design and data collection challenges encountered in data collection for the development of a model capable of detecting learned helplessness in students' behaviors while using a web application for solving linear equation. Challenges included outdated devices, unreliable internet, and logistical constraints such as limited session durations and delays in obtaining approvals. Environmental disruptions like class cancellations and curriculum gaps further complicated the process, with only 118 out of 410 students eligible and actively participating. These obstacles highlight the complexities of collecting interaction data for detecting learned helplessness in real-world, resource-constrained educational settings.

## INTRODUCTION

Learned helplessness (LH) is a psychological condition wherein repeated negative experiences lead individuals to feel powerless to influence their circumstances (Goodall & Johnston-Wilder, 2015; He, 2021; Krejtz & Nezlek, 2016). While LH can affect individuals of all ages, children are particularly vulnerable due to their limited coping skills compared to adults (Crandall et al., 2024; Goodall & Johnston-Wilder, 2015; He, 2021). LH is especially prevalent in STEM subjects, particularly mathematics, due to factors rooted in how the subject is taught, perceived, and approached. Mathematics is a foundational academic skill where achieving high marks is critical for success (Krejtz & Nezlek, 2016). It often manifest when students, after experiencing repeated academic challenges or failures come to believe they lack the ability to succeed regardless of their efforts (Fincham et al., 1989; Gürefe & Bakalım, 2018a; Hawrot & Zhou, 2024; Sorrenti et al., 2019). The frequent testing and an overemphasis on grades often amplify students' anxiety, especially among those who require more time to process information (Gürefe & Bakalım, 2018b). The abstract nature of mathematics, coupled with a lack of concrete applications or relatable examples, further exacerbates the problem. Without adequate step-by-step guidance, students may develop cumulative knowledge gaps, reinforcing feelings of helplessness (Amadi et al., 2020; Gürefe & Bakalım, 2018b). In addition, attribution of failure due to lack of ability, decrease persistence are some of the several behaviors often attributed to this (Buckley & Sullivan, 2023; Hawrot & Zhou, 2024; Hwang, 2019; Yates, 2009).

Cultural attitudes also play a significant role in shaping students' experiences with mathematics (Goodall & Johnston-Wilder, 2015). A common fixed mindset categorizes individuals as either "math people" or "non-math people," discouraging effort and persistence when challenges arise. Mathematics is often stigmatized as inherently difficult and suitable only for the intellectually gifted, fostering a sense of inevitability about failure. This perception discourages students from engaging with the subject and reinforces a fear of mistakes. In addition, parental and teacher expectations also contribute to the problem. Negative feedback, lack of encouragement, and overly critical responses can erode students' confidence and lead them to internalize failure as a reflection of personal inadequacy rather than a normal part of the learning process. Mistakes are often seen as sources of shame, prompting some students to avoid attempting problems altogether to escape the fear of being wrong.

These aspects are also true in the Philippines, a developing country with over 23 million basic education students enrolled in the 2024 - 2025 academic year (Argosino, 2024). Mathematics is widely perceived by Filipinos as a challenging subject which contributes to LH. This societal mindset fosters fear and anxiety in students even before they engage with the subject (Goodall & Johnston-Wilder, 2015). Filipino families place significant value on academic achievement, which often leads to intense pressure on children to excel in subjects like mathematics (Alampay & Garcia, 2019). Struggling students academically are frequently labeled or teased by peers (Rodriguez-Operana, 2017), exacerbating feelings of inadequacy and discouragement. Moreover, language barriers further complicate the learning process in the Philippines. Mathematics is taught in English, which is not the first language for most students, making comprehension more challenging (Bautista & Mulligan, 2010; Caraig & Quimbo, 2022). Additionally, large class sizes in many public schools hinder personalized attention from teachers, causing struggling students to go unnoticed and their difficulties to compound over time. Limited educational resources further widen learning gaps, leaving many students to navigate the subject with minimal support. These problems are reflected to previous international assessment which shows students are struggling in STEM subjects (Mullis et al., 2020; Paris, 2019).

According to studies, LH can significantly affect a student academic performance and overall well-being. If LH is not taken seriously, it can lead to detrimental effects such as avoidance of STEM fields, decreased motivation, chronic academic underachievement, and negative emotional responses (e.g., hopelessness, low self-esteem) (Kolacinski, 2003; Suárez-Pellicioni et al., 2016). Moreover, if left unaddressed, it can further lead to more severe conditions such as anxiety and depression (Cherry, 2023; Nolen, 2024). In order to address this early on, a study was conducted to create a model to detect LH using data from students' personal profile and computer interactions. This model when created, it will be integrated into the new tutoring system to make it adaptive in providing problems, hints, instructions, etc.

This study provides a descriptive account of the design and data collection challenges encountered while developing a detection model for LH in the context of the Philippines. It details technological barriers such as limited internet access and outdated devices; logistical hurdles, including school coordination, approval processes, and managing data collection in resource-constrained environments; and ethical considerations, such as securing informed consent, protecting data privacy, and ensuring equitable participation. These real-world challenges, alongside disparities in technological literacy and resource availability, shaped the development of a mobile application for teaching linear equations. By offering a practical perspective on what actually occurs in such contexts, the study contributes to the broader discourse on the considerations essential for conducting quasi-experimental research and developing technological interventions in developing countries.

## METHOD

This study adopts a descriptive case study design to explore the implementation challenges encountered during data collection for the development of a model to detect LH in mathematics. A descriptive case study is a research method that provides a detailed account of a specific instance or phenomenon within its real-life context (Edgar & Manz, 2017). In this study, the focus is on practical, real-world obstacles and solutions observed throughout the data collection process in the context of public schools in the Philippines. By documenting the dynamic experiences of implementation, the study provides valuable insights into the intersection of design, logistics, and contextual realities.

For this study, a web application named *Adaptive Sensei* (Fig. 1) was developed and used for data collection. The application was derived from prior research by Bringula et al. (2015), who initially developed a mobile-assisted learning application for mathematics. This mobile application guided students through a step-by-step process to solve linear equations. However, as the original mobile application was no longer available, the study redeveloped it based on the technical description provided in the earlier research. During the testing phase, challenges emerged, including the impracticality of distributing the mobile app to students due to the need for installation on individual smartphones. To address this, the application was redesigned as a web-based platform, enabling data collection from multiple users simultaneously and in real-time without requiring installation.

The redesigned web application was further optimized to meet the constraints of the study. With schools approving only 30-minute to 1-hour implementation sessions, the application was streamlined for efficiency and ease of use. Additional features were incorporated, such as skipping problems, new problem schemas, student profile etc. as described to previous research (Miranda & Bringula, 2023). These modifications ensured the application could collect data within the limited timeframe while addressing logistical and technological challenges identified during the testing phase. The web-based approach proved more practical and suitable to the constraints of the targe locales. During the actual data collection, out of 410 students initially considered for the study, only 118 were eligible and participated in the data collection process. Some students were excluded due to incomplete data, while others did not participate in using the web application during the designated sessions.

**Figure 1.**

*The web application utilized in the data collection process*

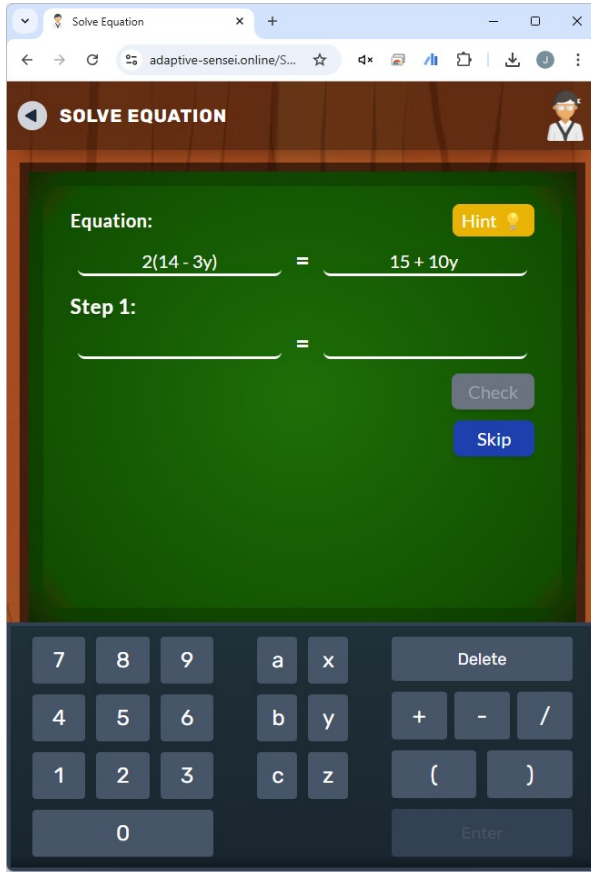

# CHALLENGES ENCOUNTERED

## Logistical and Procedural Challenges

Obtaining informed consent for this study was a complex, time-intensive process requiring coordination with multiple stakeholders, including the Department of Education's (DepEd) school division offices, school principals, head teachers, mathematics teachers, parents, and students. Before initiating data collection, the study underwent an ethics review by its ethics review committee at a top university in Manila, submitted in June 2023 and approved in January 2024. Approval was first secured from the Pampanga schools division, encompassing the City of San Fernando and Pampanga province. Despite being within the same province, separate approvals were required for each division. Following this, individual school principals had to grant permission, after which head teachers coordinated with mathematics teachers, who in turn facilitated consent from parents and students. This multi-layered process often took weeks or months for each school. This multi-layered approval process, while common in educational research, has specific implications for LH detection. The prolonged timeline between approvals and actual data collection could affect the consistency of students' mathematical experiences and their developing attitudes toward mathematics. This temporal gap might have influence how LH manifests during the data collection phase.

External factors further compounded delays. Previously, typhoons were the primary reason for class suspensions in the Philippines; however, extreme heat and even moderate rain forecasts have now become significant factors, especially in Pampanga, where much of the province is flood-prone. In some cases, even a forecast of moderate rain prompted schools to shift their learning delivery to other modes permitted by the DepEd such as modular or online learning. These frequent disruptions hindered initial contact with parents and students. Once classes resumed, schools prioritized addressing learning gaps, implementing continuity plans, and managing school events, leaving limited time to accommodate the consent process of the study. These challenges significantly prolonged the timeline for data collection and highlighted the logistical complexities of conducting research in such contexts.

## Curricular Challenges

Initial plans to target Grade 6 students required realignment to Grade 8 due to re-cent changes in the Philippine curriculum to better match students' foundational understanding of linear equations. This adjustment revealed significant implementation challenges, as half of the selected schools reported student unpreparedness stemming from disrupted prerequisite lessons and the promotion of students despite low mathematical proficiency. The gap in mathematical readiness became particularly evident when teachers reported having to reteach basic linear equation concepts to Grade 8 students, despite these being foundational topics from earlier grade levels. This situation created a critical methodological concern in distinguishing between genuine LH and simple gaps in prerequisite knowledge. While attempts to address these challenges through consultation with teachers and schools helped established suitable timing for data collection, the necessary rescheduling significantly extended the project timeline, added to logistical complexities and highlighted broader systemic challenges in curriculum implementation and student mathematical preparedness in the Philippine education system.

## Technological and Connectivity Barriers

Technological and connectivity barriers also posed challenges during the implementation of data collection for this study. Since public schools were chosen as the respondents, most were located in areas with unreliable internet connections. While the majority of students owned smartphones, many of these devices were outdated and struggled to support the web app, even though the app was designed to use minimal data and featured a simple design and graphics. Additionally, none of the schools provided free internet access for students.

Students' difficulty in using the web application presented an unexpected challenge that complicated the assessment of LH behaviors. More than half of the students struggled with basic navigation, including tasks as simple as opening the web-site. Students struggling with technical difficulties might exhibit behaviors that mimic LH (giving up, showing frustration) but are actually responses to technical barriers rather than mathematical challenges. Frequent technical support was required for various issues, including accidental app closure, navigation assistance, connectivity troubleshooting, and device malfunctions.

To address connectivity challenges, the study supplied several pocket Wi-Fi devices. However, these devices were outdated and less efficient in receiving signals com-pared to newer smartphones. Upgrading to newer Wi-Fi devices was considered, but their high cost made this option infeasible. Additionally, the location of classrooms played an important role in connectivity issues. Some classrooms were in areas with poor or dead signals, requiring the data collection team to move to multiple locations within the school premises to find better signal strength. As an additional solution, the study used up-to-date smartphones to serve as hotspots, connecting 20 – 30 students at a time to these devices. The study utilized at least six smartphones to ensure connectivity during the sessions. Only a small number of students opted to use their own mobile data, further highlighting the reliance on the provided resources. Despite these efforts, technological barriers, poor classroom signal locations, and the need for constant technical support remained significant hurdles in facilitating the data collection process.

## Factors Affecting LH Detection Accuracy

The cited barriers above also reflected the potential issues that might have significant impact on the validity of the data collected and used for developing the model for LH detection. The most prominent challenge emerged from the limited implementation windows of 30-minute to 1-hour sessions, which maybe insufficient for accurately detecting LH behaviors. These brief observation periods may have primarily captured temporary frustrations or situational responses rather than the sustained patterns of engagement typically associated with LH. Furthermore, identified curricular gaps complicated the validation process, as students' difficulties with linear equations could stem from either LH or insufficient mathematical foundations, presenting a critical challenge in distinguishing between these distinct phenomena.

The technological barriers and socio-technical context of Philippine public schools introduced additional complexity to the validation process. Student behaviors typically associated with LH such as task abandonment or avoidance, expressed frustration, or disengagement maybe triggered by technical difficulties rather than mathematical challenges. The frequent technical support interventions disrupted natural problem-solving sequences and potentially created a noise in the behavioral data. Moreover, varying levels of technological literacy among students may have influenced how LH manifested in the digital learning environment, with some students displaying behaviors that resembled mathematical LH while actually experiencing technological anxiety. The necessary adaptations utilized in this study, including problem-skipping features and streamlined application design, may have inadvertently complicated the distinction between strategic problem-solving choices and help-lessness-driven avoidance behaviors.

## CONCLUSION

The challenges encountered in this study highlighted the complexities of conducting educational research and collecting human-computer interaction (HCI) data in resource-constrained settings, particularly in developing country contexts. Logistical and procedural challenges, curricular misalignments, and technological barriers collectively contributed to delays and inefficiencies in the research process. The validation challenges in detecting LH behaviors were particularly notable, as the brief observation windows and technical disruptions complicated the distinction between genuine mathematical LH and other behavioral patterns. From a practical perspective, future researchers should prioritize designing intuitive interfaces with minimal navigation steps to reduce the need for technical support, use web-based platforms with low data requirements, and leverage flexible connectivity solutions like portable hotspots to improve reliability. Understanding the socio-cultural and educational landscape remains critical, including aligning study timelines with school schedules and addressing curriculum gaps, as well as establishing early communication with stakeholders to minimize logistical hurdles.

To address the challenges identified in this study, future implementations require both practical and methodological refinements. The integration of baseline techno-logical comfort assessments, mathematical prerequisite testing, and extended observation periods would strengthen the validity of LH detection. The development of more nuanced behavioral markers capable of distinguishing between technology-induced and mathematics-induced helplessness patterns is also crucial for enhancing and improving the accuracy of the model. Contingency plans, such as alternative data collection locations and accounting for rescheduling, can further mitigate disruptions. This study ultimately demonstrates the need for context-sensitive strategies that not only simplify user engagement and enhance system adaptability but also integrate affordable and scalable technological solutions. Such comprehensive approaches are essential for advancing adaptive intelligent systems like the creation of a model detector for LH in under-resourced contexts, particularly where technological, curricular, and logistical challenges intersect.

## PRACTICAL IMPLICATIONS

The complexity of the approval process and its impact on research timelines underscores the importance of building substantial buffer time for administrative procedures when planning to conduct studies in similar contexts. Researchers should anticipate that obtaining permissions from multiple stakeholders, such as the education department, offices, school principals, teachers, and parents, could take several months. This has direct implications for research design, particularly for time-sensitive studies or those with strict funding deadlines. Furthermore, new and emerging threats, like the vulnerability to climate-related disruptions, highlighted the need for more flexible and adaptive research methodologies. The finding that even mod-erate rain forecasts can trigger shifts to alternative learning modes suggests that re-searchers need to develop contingency plans that can accommodate both in-person and possibly remote data collection. This could include developing hybrid research instruments that work across different delivery modes whenever possible.

The universal assumption that students are universally "digital natives" was challenged by the findings, particularly in the context of Philippine students, as many struggled with basic digital navigation. These findings have broader implications for educational policy and practice in developing countries. This study suggests that digital learning initiatives need to consider not just device access but also internet infra-structure, technical support capacity, and students' digital literacy levels. The challenges encountered also emphasized the importance of conducting realistic assessments of students' prerequisite knowledge and skills, especially in mathematics, where concepts build upon each other sequentially. Policymakers and educators should recognize that the mere provision of digital devices does not automatically translate to effective digital learning. Instead, a comprehensive approach that ad-dresses infrastructure, support, and digital literacy is necessary to ensure that students can fully benefit from digital learning initiatives.